# Suppression of indirect exchange and symmetry breaking in antiferromagnetic metal with dynamic charge stripes

K. Krasikov[1], V. Glushkov[1], S. Demishev[1,2], A. Khoroshilov[1], A. Bogach[1], V. Voronov[1], N. Shitsevalova[3], V. Filipov[3], S. Gabáni[4], K. Flachbart[4], K. Siemensmeyer[5], N. Sluchanko[1]

[1]Prokhorov General Physics Institute of Russian Academy of Sciences, Vavilov str. 38, Moscow 119991, Russia

[2]National Research University Higher School of Economics, Myasnitskaya str., 20, Moscow 101000, Russia

[3]Frantsevich Institute for Problems of Materials Science, National Academy of Sciences of Ukraine, 03680 Kyiv, Ukraine

[4]Institute of Experimental Physics, Slovak Academy of Sciences, Watsonova 47, 04001 Košice, Slovakia

[5]Helmholtz Zentrum Berlin, Hahn Meitner Platz 1, D 14089, Berlin

**Abstract.**

Precise angle-resolved magnetoresistance (ARM) measurements are applied to reveal the origin for the lowering of symmetry in electron transport and the emergence of a huge number of magnetic phases in the ground state of antiferromagnetic metal $HoB_{12}$ with *fcc* crystal structure. By analyzing of the polar $H$-$\theta$-$\varphi$ magnetic phase diagrams of this compound reconstructed from the experimental ARM data we argue that non-equilibrium electron density oscillations (dynamic charge stripes) are responsible for the suppression of the indirect RKKY exchange along <110> directions between the nearest neighboring magnetic moments of $Ho^{3+}$ ions in this strongly correlated electron system.

The concurrence between different simultaneously active charge, spin, lattice, and orbital interactions is believed to play a key role in the formation of a rich variety of magnetic structures in phase diagrams of strongly correlated electron systems (SCES) [1]. A well-known example is manganites with colossal magnetoresistance that demonstrate an enormous number of different states including ferro- and antiferromagnetic (AF) long-range ordering, magnetic short-range correlations, charge and spin density waves, and diversity of structural transitions [1-3]. In the family of cuprates, the competition of high temperature superconductivity with AF ordering is complemented by spin and charge ordering, pseudogap states and the unusual "strange metal" phase with non-Fermi-liquid behavior at intermediate temperatures [4-6]. Very complicated phase diagrams have been also found in ruthenates [7], organic charge-transfer salts [8], superconducting iron-based pnictides and chalcogenides [9,10] and in Ce-based heavy fermion metals [11], where electron nematic phases were detected. The contention of various states leads often to nanoscale phase separation and spatially inhomogeneous phases that is extremely useful for practical applications due to giant responses to weak enough external perturbations. Unfortunately, the common features of the effects arising in most of these SCES cannot be easily analyzed since these compounds possess a complex chemical composition with a low symmetry of the crystal structure.

It was recently demonstrated that $Ho_xLu_{1-x}B_{12}$ dodecaborides with a face-centered cubic (*fcc*) crystal structure may be treated as model SCES, which show electronic phase separation (dynamic charge stripes along the <110> axis, Fig.1a) in combination with dynamic Jahn-Teller instability and a frustrated AF ground state with Néel temperature $T_N \leq 7.4$ K [12-15] . In these compounds, the competition between Ruderman-Kittel-Kasuya-Yosida (RKKY) oscillations of the electron spin density (indirect exchange interaction) and quantum oscillations of the electron density (dynamic charge stripes) is shown to result in the extremely complicated and anisotropic magnetic phase diagram, which have a form of maltese cross with a huge number of various magnetic phases [14-16]. Since the arrangement of magnetic phases depends dramatically both on the magnitude and direction of the external magnetic field the high precision studies of charge transport was performed here to reconstruct the three-dimensional (*3D*) *H-φ-θ* magnetic phase diagram of $HoB_{12}$. As a result, we argue that non-equilibrium electron density oscillations (stripes) seem to be responsible for the suppression of the indirect exchange interaction along <110> directions and lead to symmetry breaking in this strongly correlated electron system.

The detailed study of heat capacity, magnetization and magnetoresistance was performed for high-quality single-domain and isotopically enriched ($^{11}$B) single crystals of $Ho^{11}B_{12}$ ($T_N \approx$ 7.4 K). The related experimental details are given in Supplementary materials [17]). The *H-T* magnetic phase diagrams were reconstructed for three principal directions ***H*** ∥ [100], ***H*** ∥ [110] and ***H*** ∥ [111] displaying numerous magnetic phases and phase transitions (Figs.1c-1e). It is worth noting that in contrast to the nearly isotropic AF-P transition (P- paramagnetic state) in the *H-T* diagram, the location of phase boundaries inside the AF phase is considerably different for various ***H*** directions (Figs.1c-1e). The results of our precise measurements allowed us to refine the *H-T* diagrams detected previously in [16,18,19]. Besides, we show below that despite high symmetry of the crystal lattice *the magnetic phases for the principal field directions* are separated by the radial phase boundaries and, hence, these *are completely different,* and that only one AF phase (marked as I in Fig.1) exists for any magnetic field orientation.

To clarify the location of phase boundaries in the AF state in the external magnetic field up to 80 kOe, the angle-resolved magnetoresistance (ARM) was measured at fixed temperature T = 2.1 K (see vertical dotted line in Figs.1c-1e) in four experiments with sample rotating around excitation current directions (i) ***I***∥[100], (ii) ***I***∥[1$\bar{1}$0], (iii) ***I***∥[111] and (iv) ***I***∥[112]. The details can be found in Supplementary materials (see example of experimental data (i) with rotating around ***I***∥[100] in Figs. 3-4, [17]). Results of these experiments may be summarized as follows. Firstly, only slight cosine-like $\rho(\varphi) \sim \cos(2\varphi)$ modulation is detected when the magnitude of H does not exceed ~20 kOe (the phase boundary position, which restricts the "I" phase, see Figs.1c-1e). Secondly, the increase the magnetic field above ~25 kOe leads two additional features (sharp peaks in the neighborhood of <110> directions and "horns" that are symmetrical to the <100> axis) appear and become broader with ***H*** increase. Thirdly, the cosine-like behavior is restored above the Neel field ($H_N$ ~ 76 kOe) with two additional features, different from the low-field data: (*a*) the phase is shifted by 45 degrees and (*b*) a narrow local minimum of tiny amplitude appears near the <001> direction.

It is worth noting that abrupt anomalies in the AF phase observed in the wide neighborhood of <100> directions may be associated with magnetic phase transitions between states with different magnetic order [13,14]. Besides, the unusual behavior of sharp peaks in the vicinity of <110> directions in $HoB_{12}$ is suggested to be induced by the dynamic charge stripes along <110> axis (see Fig.1a).

The same features can be easily resolved on the magnetic field dependences of resistivity measured at the same temperature T = 2.1 K for the set of different ***H*** directions. For example, the data for ***I***∥[100] presented in supplementary materials unambiguously evidence that (1) varying of the angle $\varphi = (\bm{n}, \bm{H})$ between the normal to the sample surface ***n***∥[010] and the applied magnetic field ***H*** allows detecting the set of anomalies on the $\rho(H, \varphi_0)$ curves, which may be attributed to orientation magnetic transitions. We showed that (2) positive magnetoresistance dominates in the AF phase up to the sharp drop just before the Neel field. It is important that a wide hump appears in the range of H=22-35kOe (in the proximity (Δφ=2-3°) of [110] direction (when a magnetic field is applied along the dynamic charge stripes). Similar data sets have been collected in our experiments with sample rotation around current directions (ii) *I*‖[1$\bar{1}$0], (iii) *I*‖[111] and (iv) *I*‖[112] (see [17], Figs. 4-5). The total review of both the magnetic phases and phase transitions inside the AF state of Ho$^{11}$B$_{12}$ on the basis of our experiments is provided below.

The extracted ARM data for all (i)-(iv) sample rotation experiments can be easily overlooked in the polar presentation $\Delta\rho/\rho = f(H, \varphi)$. Figure 2 shows the polar plot of the MR results at T = 2.1 K obtained in the experiments (i) (panel (a), rotation axes *I*‖[100]) and (ii) (panel (b), *I*‖[1$\bar{1}$0]) together with their projection on the (100) and (1-10) planes (panels (c) and (d), respectively. Two other sets of the ARM data and the diagrams for (iii) *I*‖[111] and (iv) *I*‖[112] rotation axes are presented in [17]. The phase boundaries on the *H-φ* plane diagrams (open symbols in Fig. 2 c-d) are related to the sharp features detected on the Δρ/ρ angular dependencies and MR derivatives dρ/dH [17]. It is worth noting that the rotation around the *I*‖[1-10] axis enables us to obtain in one experiment the transverse MR variation when the direction of magnetic field changes step-by-step between three principal axes in the *fcc* lattice (***H***‖[001], ***H***‖[1-10] and ***H***‖[1-11]). Several regions corresponding to the phases with different magnetic ordering are easily resolved in Figs.2b and 2d. The obtained *H-φ* diagram (Fig.2d) appears to be crucial for further understanding of the structure of the polar *H-φ-θ* phase diagrams. Similar "Maltese Cross" anisotropy in the (1-10) plane was found recently [15] in the related Ho$_{0.8}$Lu$_{0.2}$B$_{12}$ system. Such a kind of MR anisotropic is constructed from four principal regions that can be easily distinguished in Fig. 2d: (1) the central circular-like region with a radius of H ~ 20kOe; (2) the area in the wide neighborhood of the [100] direction where the maximum MR values are achieved; (3) the region with the minimum MR values around the [111] direction; (4) the zone with positive MR that is symmetrical concerning the direction of dynamic charge stripes ***H***‖[110].

Note, that the edges of these plane *H-φ* diagrams (Figs. 2c and 2d and [17]) provide the important information related to the arrangement of four different sections of the *cylindrical H-φ-θ (3D) magnetic phase diagram* in the cases when ***H*** vector lies in planes to be perpendicular to the measuring current axes. Figure 3a shows the combined *H-φ* planes for measuring current directions *I*‖[100] and *I*‖[1$\bar{1}$0] (see Figs.2c-2d) for Ho$^{11}$B$_{12}$ at T = 2.1 K and presents also schematic views of the main regions (phases I, III, IV, and VIII in Figs.2c-2d) in the *3D H-φ-θ* space constructed with the help of two more sets of experimental data ((iii) *I*‖[111] and (iv) *I*‖[112], see [17]). Since these four *H-φ* planes for various exciting current directions are obtained for the different samples cut from one single crystal (***n***‖[110] plate) of Ho$^{11}$B$_{12}$, we can compare the behavior of MR rather than the absolute values of the MR amplitude. However, it can be seen that the phase boundaries obtained from these experiments with different rotational axes coincide very well with each other (see also [17]). One more schematic view (the projection on the spherical surface) of three basic high-field phases (III, IV, and VIII) is shown in Fig.3b. It is worth noting that these three segments corresponding to three different types of magnetic phases (III, IV, and VIII, see Figs.2 and 3) in combination with the low-field spherical area (phase I, Figs.2c, 2d and 3a) fill almost completely the space inside the AF region in the *3D H-φ-θ* phase diagram. The dashed lines in Fig.3b indicate the angular $\varphi = (\boldsymbol{n}, \boldsymbol{H})$ trajectories in these four experiments: (i) A-B-A corresponds to the rotation from ***H***‖[010] to ***H***‖[001] (see Fig.2c); (ii) A-C-B corresponds to the rotation from ***H***‖[001] to ***H***‖[110] (see Fig.2d); (iii) B-c-B corresponds to the rotation from ***H***‖[011] to ***H***‖[110] [17] and (iv) B-a-b-C corresponds to the rotation from ***H***‖[011] to ***H***‖[111] [17]. Thus, it is obvious that despite the partial similarity of the borders on the *H-T* phase diagrams of Ho$^{11}$B$_{12}$ for various directions of the external magnetic field (Fig.1c-e), the set of phases in three principal directions is almost completely different. Such a pronounced anisotropy of charge carriers scattering and of the magnetic phase diagram should be associated, from one side, with the interaction of the external magnetic field with the dynamic charge stripe structure in the Ho$^{11}$B$_{12}$ matrix. Besides, if consider the stripes as the main factor to be responsible for a strong renormalization of the indirect RKKY exchange interaction one needs to expect a huge suppression of the nearest neighbor exchange ($J_1$) in this model AF metal. Indeed, the long-range RKKY magnetic interaction between the localized magnetic moments of *d*-, or *f*- orbitals in metals is transmitted by the spin density oscillations of conduction electrons (see fig.1b), and these are suppressed dramatically in the presence of the dynamic charge stripes (fast quantum vibrations of the non-equilibrium charge density with a frequency ~200 GHz [20]) directed along <110> axes. Thus, the stripe direction just corresponds to the location of neighboring magnetic ions (figs.1a) and it destroys totally the nearest neighbor RKKY interaction (fig.1).

To support this ARM experimental finding an independent estimate of the nearest neighbor ($J_1$) and next-nearest neighbor ($J_2$) interactions is undertaken here. With this purpose, the magnon dispersion data in HoB$_{12}$ [19] are compared with the classical Monte Carlo (MC) simulation. The dispersion is derived from the MC generated structure by numerical integration of the classical equation of motion for a moment $\vec{M}$ in a field $\vec{B}$, $\dot{\vec{M}} = -\gamma \vec{M} \times \vec{B}$, where $\gamma$ is the gyromagnetic ratio and field $\vec{B}$ is the local exchange field obtained from the MC simulation. This first gives the time dependence of the moments which then is converted to the dispersion by a Fourier transform (see [21] for details). The dispersion has been calculated for a system of $4 \ast 16^3$ spins cooled down from the paramagnetic state to an MC temperature of T=0.025 in 5000 steps. At each temperature, the structure was annealed. The nearest neighbor interaction $J_1$ was varied between -5 and 5 in steps of 0.5 and $J_2 = 3$ was kept fixed. In Fig. 3c we show the measured dispersion along the (½, k,k) direction [19] together with the simulation for $J_1 = 0$. For this value, the flat mode found in experiments [19] is reproduced very well and it is not the case for any finite $J_1$. Deviations between simulation and experiment at the commensurate positions, e.g. at (½, ½, ½), are due to domain walls in the MC generated structure, leading to the strong streaks there. Note also, that the simulation does not account in detail crystal field anisotropy, which causes the energy gap in HoB$_{12}$.

Summarizing up, the model SCES Ho$^{11}$B$_{12}$ with incommensurate AF structure, cooperative Jahn-Teller instability of boron network and dynamic charge stripes was studied in detail by ARM measurements at liquid helium temperature. In this non-equilibrium AF metal, a strongly anisotropic polar *H*-*φ*-*θ* magnetic phase diagrams were reconstructed for the first time. These are shown to consist of four basic sectors: spherical low-field region and three different complicated cone-shaped segments detected in the vicinity of the principal directions: (a) along (**H**||[110]) and (b) transverse to (**H**||[001]) the dynamic charge stripes, and (c) parallel to the axis of magnetic structure (**H**||[111]) in the *fcc* lattice. We argue that *the strong anisotropy* both of the phase diagram and the charge transport *is a fingerprint of the electron instability* related to the formation of a filamentary structure (fluctuating charge carrier channels along [110] direction) of non-equilibrium electrons. As a result, *the indirect* RKKY *exchange interaction* between the nearest neighbored magnetic ions located at the distance of ~5.3 Å in the <110> directions *is dramatically destroyed providing magnetic symmetry lowering* and leading to field-angular phase diagrams with a huge number of magnetic phases and phase transitions.

**Acknowledgments**
This work was supported by the Russian Science Foundation, Project No. 17-12-01426 and was performed using the equipment of the Shared Facility Center for Studies of HTS and Other Strongly Correlated Materials, Lebedev Physical Institute, the Russian Academy of Sciences, and the Center of Excellence, Slovak Academy of Sciences. The work of K.F. and S.G. is supported by the Slovak agencies APVV (Grant No. 17-0020) and DAAD-SAS (Grant No. 57452699). The authors are grateful to V. Krasnorussky for experimental assistance.

**References**

[1]  E. Dagotto, Complexity in strongly correlated electronic systems, Science **309**, 257 (2005). 10.1126/science.1107559
[2]  J. Mitchell, D. Argyriou, A. Berger, K. Gray, R. Osborn, and U. Welp,  (ACS Publications, 2001).
[3]  E. Dagotto, Nanoscale Phase Separation and Colossal Magnetoresistance (Springer, Berlin, 2003).
[4]  B. Keimer, S. A. Kivelson, M. R. Norman, S. Uchida, and J. Zaanen, From quantum matter to high-temperature superconductivity in copper oxides, Nature **518**, 179 (2015).
[5]  E. Berg, E. Fradkin, S. A. Kivelson, and J. M. Tranquada, Striped superconductors: how spin, charge and superconducting orders intertwine in the cuprates, New J Phys **11**, 115004 (2009).
[6]  S. Sachdev, and Keimer, B, Quantum criticality. Phys. Today **64**, 29 (2011).
[7]  S. Nakatsuji, V. Dobrosavljević, D. Tanasković, M. Minakata, H. Fukazawa, and Y. Maeno, Mechanism of hopping transport in disordered Mott insulators, Phys Rev Lett **93**, 146401 (2004).
[8]  T. Sasaki, N. Yoneyama, A. Matsuyama, and N. Kobayashi, Magnetic and electronic phase diagram and superconductivity in the organic superconductors κ−(ET) 2 X, Phys Rev B **65**, 060505 (2002).
[9]  R. M. Fernandes, A. V. Chubukov, and J. Schmalian, What drives nematic order in iron-based superconductors?, Nat Phys **10**, 97 (2014).
[10]  J. Lee, F. Schmitt, R. Moore, S. Johnston, Y.-T. Cui, W. Li, M. Yi, Z. Liu, M. Hashimoto, and Y. Zhang, Interfacial mode coupling as the origin of the enhancement of T c in FeSe films on SrTiO 3, Nature **515**, 245 (2014).
[11]  S. Demishev, V. Krasnorussky, A. Bogach, V. Voronov, N. Y. Shitsevalova, V. Filipov, V. Glushkov, and N. Sluchanko, Electron nematic effect induced by magnetic field in antiferroquadrupole phase of CeB 6, Sci Rep-Uk **7**, 1 (2017).
[12]  N. Sluchanko, A. Bogach, N. Bolotina, V. Glushkov, S. Demishev, A. Dudka, V. Krasnorussky, O. Khrykina, K. Krasikov, and V. Mironov, Rattling mode and symmetry lowering resulting from the instability of the B 12 molecule in LuB 12, Phys Rev B **97**, 035150 (2018).
[13]  N. B. Bolotina, A. P. Dudka, O. N. Khrykina, V. N. Krasnorussky, N. Y. Shitsevalova, V. B. Filipov, and N. E. Sluchanko, The lower symmetry electron-density distribution and the charge transport anisotropy in cubic dodecaboride LuB12, Journal of Physics: Condensed Matter **30**, 265402 (2018).
[14]  N. Sluchanko, A. Khoroshilov, V. Krasnorussky, A. Bogach, V. Glushkov, S. Demishev, K. Krasikov, N. Y. Shitsevalova, and V. Filippov, Magnetic Phase Transitions and the Anisotropy of Charge Carrier Scattering in Antiferromagnetic Metal Ho 0.5 Lu 0.5 B 12 with Dynamic Charge Stripes, Bulletin of the Russian Academy of Sciences: Physics **83**, 853 (2019).
[15]  A. L. Khoroshilov, V. N. Krasnorussky, K. M. Krasikov, A. V. Bogach, V. V. Glushkov, S. V. Demishev, N. A. Samarin, V. V. Voronov, N. Y. Shitsevalova, V. B. Filipov, S. Gabani, K. Flachbart, K. Siemensmeyer, S. Y. Gavrilkin, and N. E. Sluchanko, Maltese cross anisotropy in Ho0.8Lu0.2B12 antiferromagnetic metal with dynamic charge stripes, Phys. Rev. B **99**, 174430 (2019). 10.1103/Physrevb.99.174430
[16]  A. Khoroshilov, V. Krasnorussky, A. Bogach, V. Glushkov, S. Demishev, A. Levchenko, N. Shitsevalova, V. Filipov, S. Gabáni, and K. Flachbart, Anisotropy of Magnetoresistance in HoB12, Acta Physica Polonica, A. **131**, 976 (2017). 10.12693/APhysPolA.131.976
[17]  See Supplemental Material at [link] for more information: Experimental details. Temperature dependencies of the specific heat for **H**||[100], **H**||[110] and **H**||[111] (Fig.S1), magnetization dependencies (Fig.S2a-c) and dM/dH derivatives (Fig.S2d-f) versus magnetic field for **H**||[100], **H**||[110] and **H**||[111] directions were used to refine H-T phase diagram

(Fig.1c-e). Example of experimental ARM data obtained at T=2.1 K - angular (Fig.S3) and magnetic field (Fig.S4) dependencies of MR. Magnetic field derivatives of MR (Fig.S5) were used to obtain angular phase boundaries on the H-φ phase diagram (Fig.2c). Magnetoresistance in cylindric coordinates and its projections onto (111) (Fig.S6a) and (112) (Fig.S6b) planes, rotation was performed at T=2.1 K around **I**∥[111] and **I**∥[112] axes respectively.

[18] A. Kohout, I. Batko, A. Czopnik, K. Flachbart, S. Matas, M. Meissner, Y. Paderno, N. Shitsevalova, and K. Siemensmeyer, Phase diagram and magnetic structure investigation of the fcc antiferromagnet HoB12, Phys Rev B **70**, 224416 (2004). 10.1103/Physrevb.70.224416

[19] K. Siemensmeyer, K. Habicht, T. Lonkai, S. Mat'as, S. Gabani, N. Shitsevalova, E. Wulf, and K. Flachbart, Magnetic properties of the frustrated fcc - Antiferromagnet HoB12 above and below T (N), J. Low. Temp. Phys. **146**, 581 (2007). 10.1007/s10909-006-9287-4

[20] N. Sluchanko, A. Azarevich, A. Bogach, N. Bolotina, V. Glushkov, S. Demishev, A. Dudka, O. Khrykina, V. Filipov, and N. Y. Shitsevalova, Observation of dynamic charge stripes in Tm0. 19Yb0. 81B12 at the metal–insulator transition, Journal of Physics: Condensed Matter **31**, 065604 (2018).

[21] T. Huberman, D. Tennant, R. Cowley, R. Coldea, and C. Frost, A study of the quantum classical crossover in the spin dynamics of the 2D S= 5/2 antiferromagnet Rb2MnF4: neutron scattering, computer simulations and analytic theories, Journal of Statistical Mechanics: Theory and Experiment, P05017 (2008).

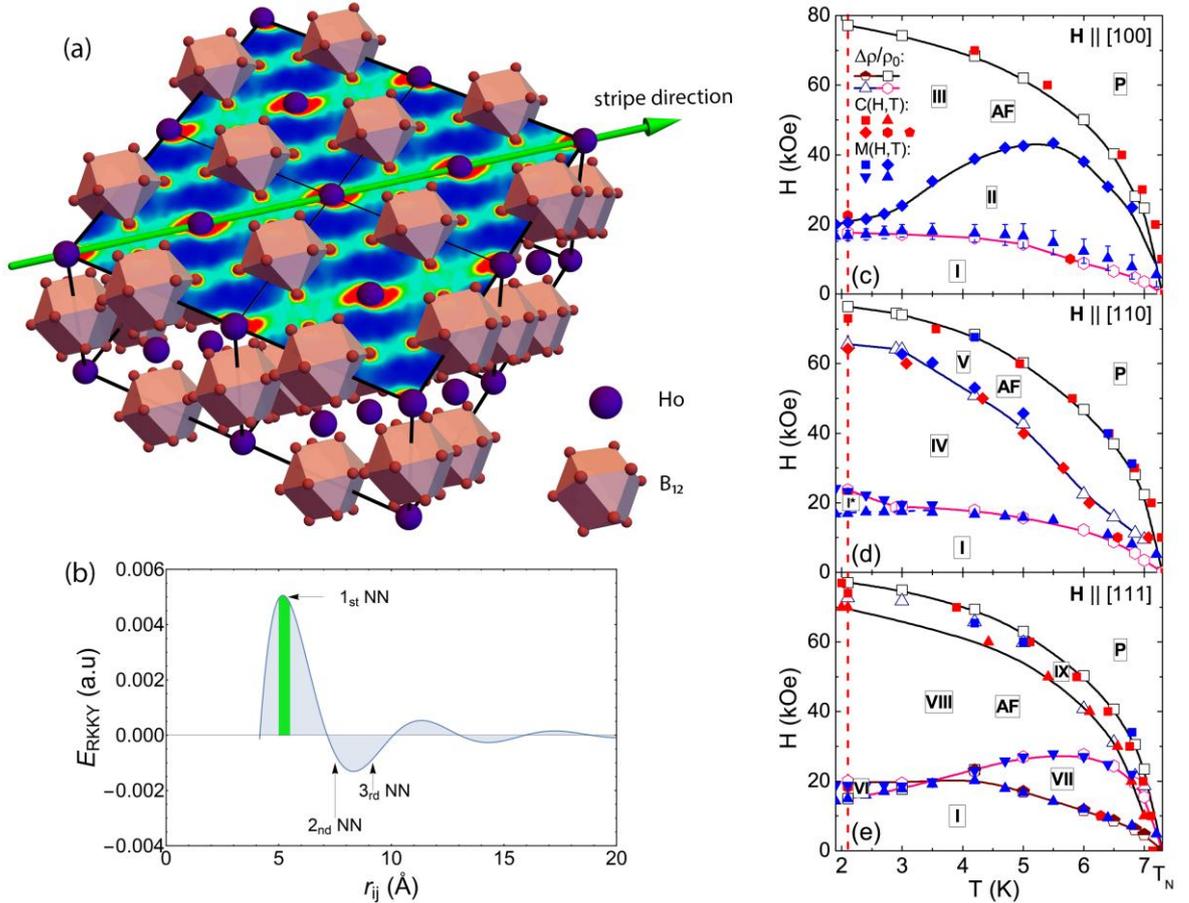

**Fig.1.** (a) Crystal structure of $R$B$_{12}$. The color plane shows the distribution of the electron density in the dynamic charge stripes (green bands) along with [110] direction [12]. (b) RKKY exchange interaction function in $R$B$_{12}$. The vertical green band marks the distance where the dynamic charge stripes destroy the indirect magnetic exchange. Arrows show the position of first (1$_{st}$ NN), second (2$_{st}$ NN) and third (3$_{st}$ NN) nearest neighbored magnetic ions in the *fcc* lattice. (c)-(e) *H-T* magnetic phase diagrams of Ho$^{11}$B$_{12}$ obtained from magnetoresistance $\Delta\rho/\rho_0$, heat capacity $C(T, H)$ and magnetization $M(T, H)$ measurements (see legend in panel c) for three

principal directions $H\|[100]$, $H\|[110]$ and $H\|[111]$ (panels c, d, and e respectively). Roman numerals indicate different AF phases, P – paramagnetic state. The red dotted vertical line indicates the temperature T = 2.1 K where the 3D ($H$-$\varphi$-$\theta$) diagram was measured.

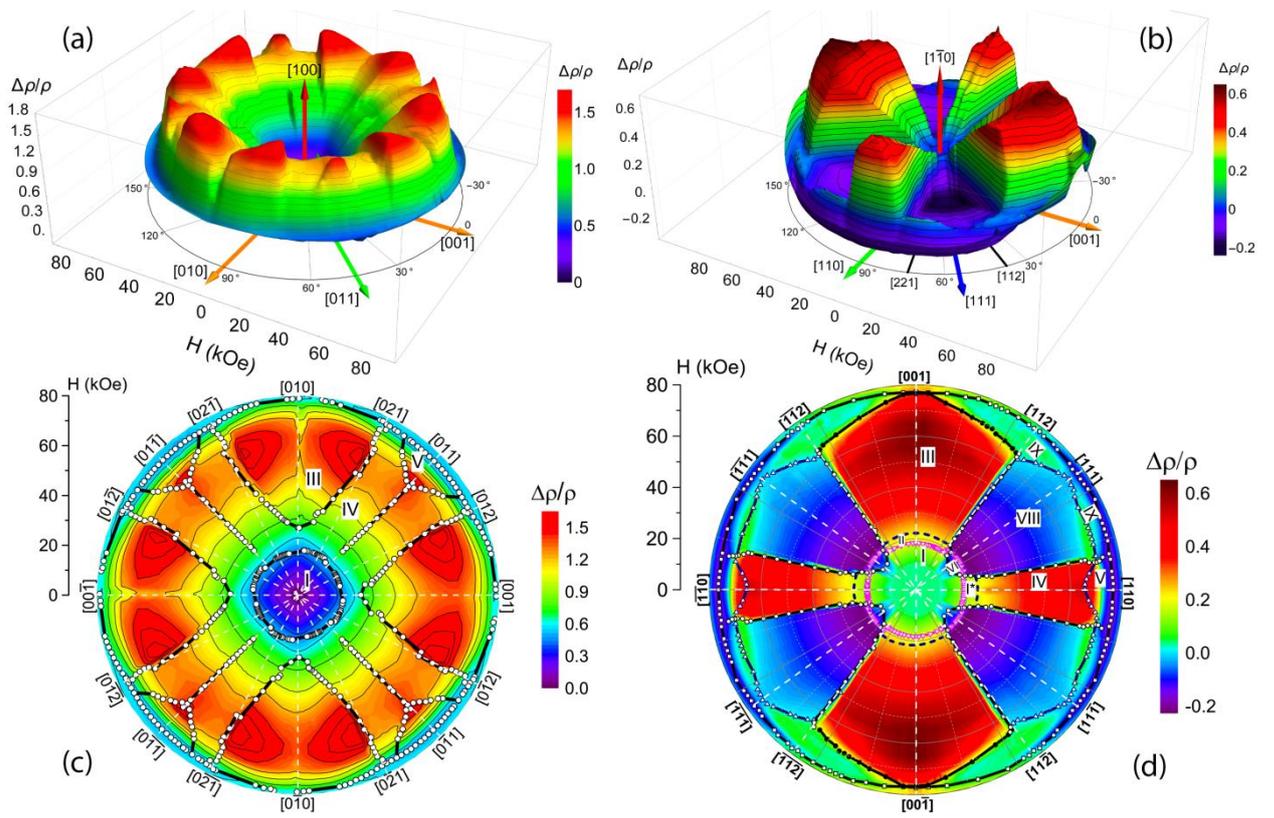

**Fig.2.** Magnetoresistance $\Delta\rho/\rho = f(H, \varphi)$ of Ho$^{11}$B$_{12}$ in cylindrical coordinates for current directions **I**∥[100] (a) and **I**∥[1$\bar{1}$0] (b) and its projection onto (100) and (110) surfaces (c) and (d) respectively at T = 2.1 K. Roman numerals show different magnetic phases in the AF state.

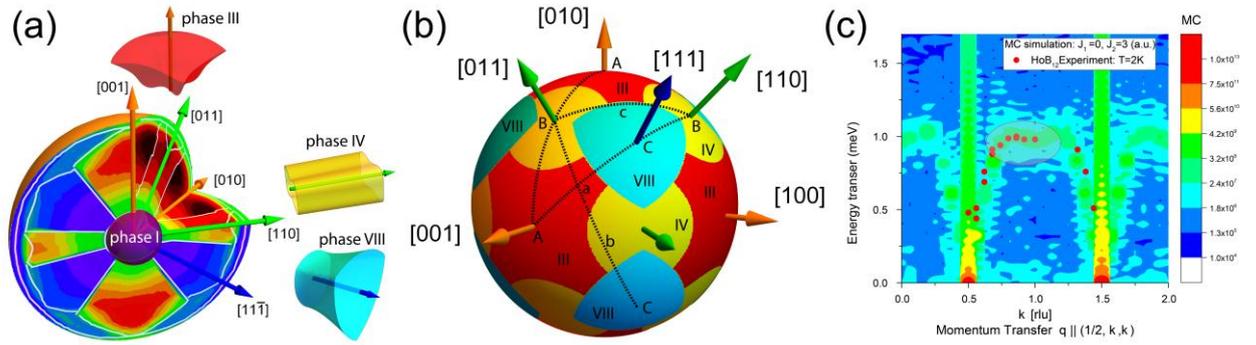

**Fig.3.** (a) *H-φ* phase diagrams of HoB$_{12}$ for current directions **I**||[100] and **I**||[1$\bar{1}$0] combined into one 3D *H-φ-θ* phase diagram. Purple, cyan, yellow and red shapes show four key segments (phases I, VIII, IV, and III, correspondingly) in the center and in the vicinity of principal axes that determine the anisotropy of the phase diagram. (b) Schematic view of the projection on the spherical surface *H*~ 60 kOe of three basic high-field phases (III, IV and VIII) in the *H-φ-θ* phase diagram. The dashed lines in panel (b) indicate the trajectories of angle *φ* variation in four experiments with sample rotation (see text for details). (c) The measured magnon dispersion of HoB$_{12}$ [19] in comparison with a dispersion obtained from a Monte Carlo simulation where the time dependence is simulated by numerical integration of the equation of motion [21]. The energy of the simulated results is scaled to match the experimental graph.

# Supplementary materials

## Experimental details

The high quality $Ho^{11}B_{12}$ single crystals were grown by vertical crucible-free inductive floating zone melting in an inert gas atmosphere on a setup described in detail in [1]. Resistivity measurements were performed using a standard DC 4-probe technique with a commutation of measuring current. Sample rotation around the **I** direction in an external magnetic field up to 80kOe was made on the original inset that can rotate the crystal over the range 360° under control from a stepper motor. High accuracy of temperature stabilization (ΔT<0.002K at 2.1K) was achieved by using the commercial temperature controller TC 1.5/300 (Cryotel Ltd.) in combination with a thermometer CERNOX 1050 (Lake Shore Cryotronics, Inc.). To refine the phase boundaries on the H-T diagram, we also analyzed the temperature and magnetic field dependences of the magnetization and specific heat obtained for a magnetic field directions **H**∥[100], **H**∥[110] and **H**∥[111], on the MPMS-5 and PPMS-9 (Quantum Design) installations respectively.

## Heat capacity and magnetization measurements

Temperature dependencies of heat capacity for various amplitudes and directions of the external magnetic field are presented in fig.1. Different phase transitions (marked as arrows) were observed: a huge C(T) drop corresponds to AF-P transition, and several peaks and kinks that were observed in the AF-phase show the presence of orientation magnetic transitions. It is easy to see that AF–P transition shifts with increasing *H* downward, demonstrating the typical tendency to suppress the AF magnetic order state by a strong external magnetic field.

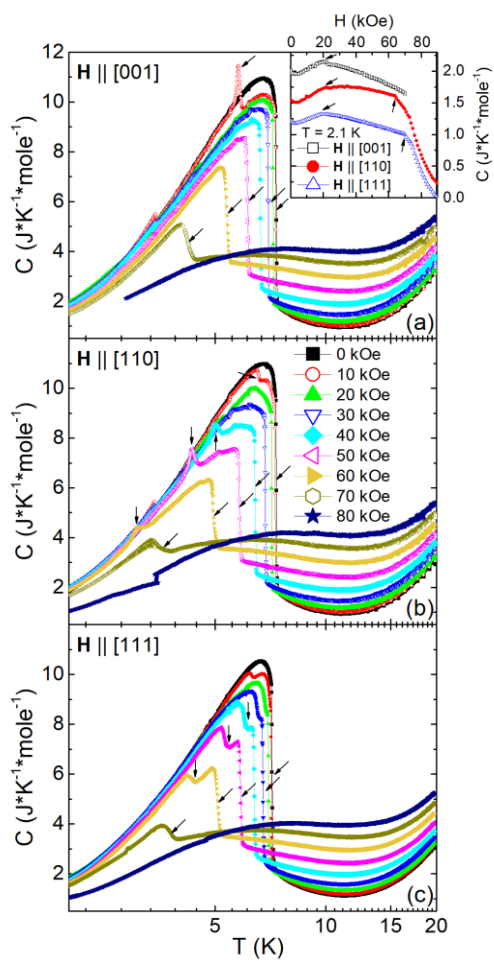

Fig.1 Temperature dependencies of Ho$^{11}$B$_{12}$ specific heat for **H**||[100], **H**||[110] and **H**||[111] – (a),(b) and (c) respectively. The inset on panel (a) shows magnetic field dependencies of the heat capacity at T=2.1K for the magnetic field directed along principal axes. Arrows show orientation magnetic phase transitions.

An additional refinement of the phase boundaries was made using dependencies of the magnetization [panels (a)-(c)] and magnetic susceptibility $dM/dH = f(H, T_0)$ obtained by numerical differentiation [panels (d)–(f)] for three principal field directions up to 50 kOe in the temperature range of 1.9–7.2K. As can be seen from Figs.2(a-c), the AF-phase transition only slightly changes the absolute magnetization values in the entire range of $H$. At the same time, details of anomalies on $dM/dH = f(H)$ curves associated with magnetic phase transitions in the range of 5–30 kOe are noticeably different for various directions of the external magnetic field. Sharp transitions in the vicinity of H≈18kOe were observed for all **H** directions, in contrary with transitions at range H>20kOe for **H**||[001] and **H**||[111], which positions significantly vary from each other.

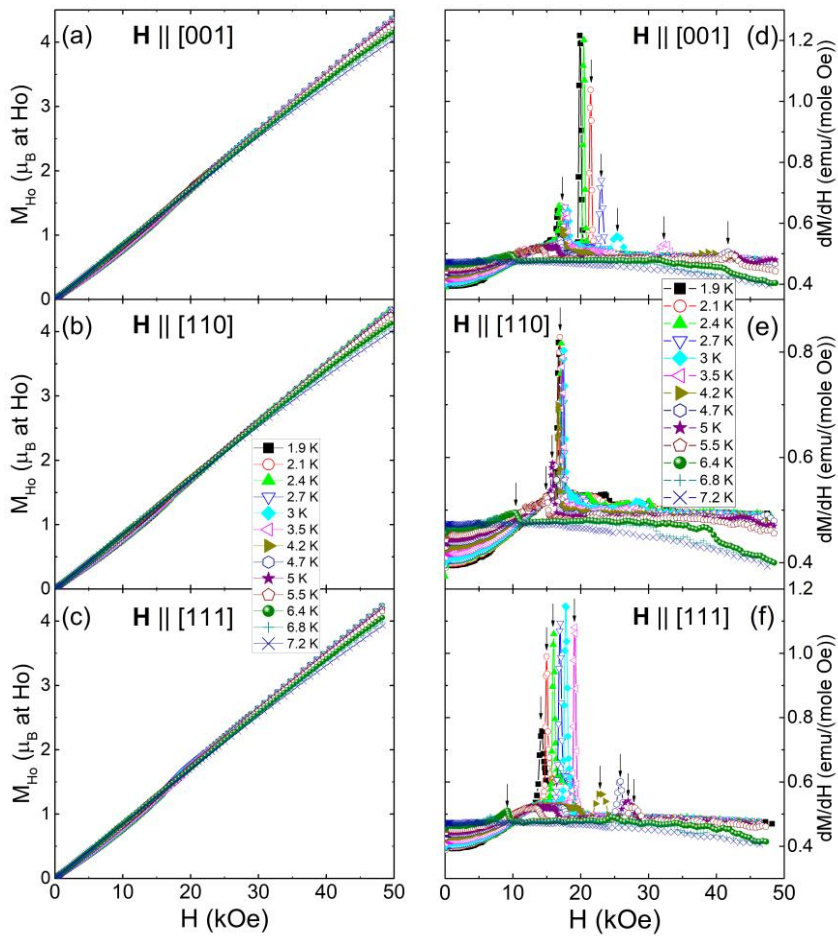

Fig.2 Magnetic field dependencies of magnetization (a-c) and dM/dH derivatives (d-f) for various temperatures and different directions of the external magnetic field **H**. Arrows show orientation magnetic phase transitions.

**Angular and field dependences of resistivity**

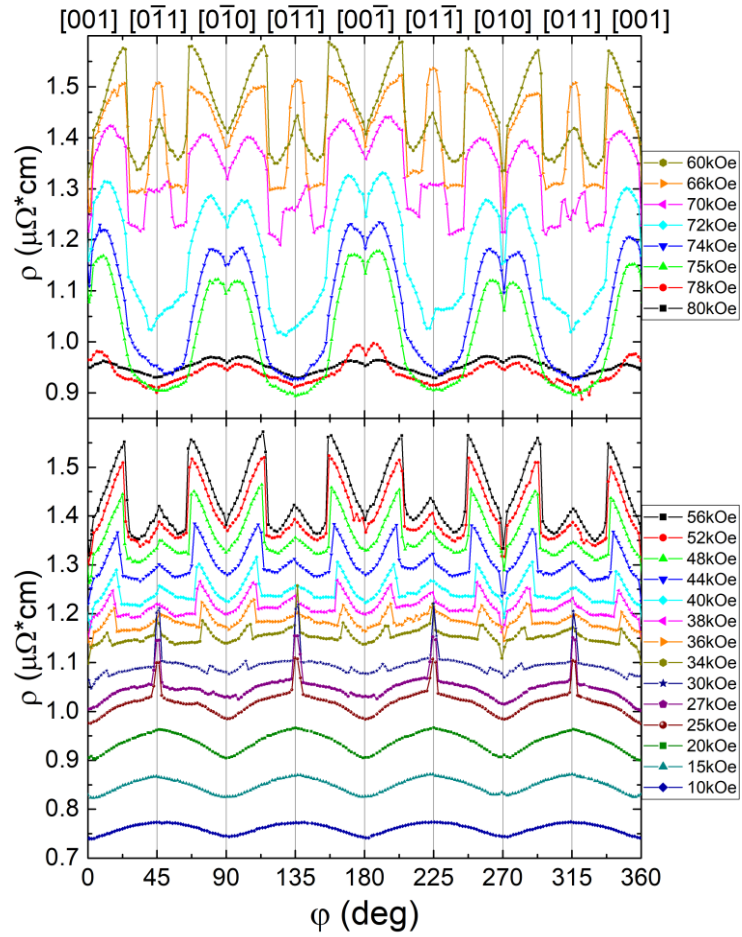

**Fig.3.** Angular dependencies of Ho$^{11}$B$_{12}$ resistivity for the various magnetic field at $T = 2.1$ K. The rotation was performed around the axis $I\|[100]$. Vertical lines show positions when the magnetic field $H$ is aligned with the principal axes.

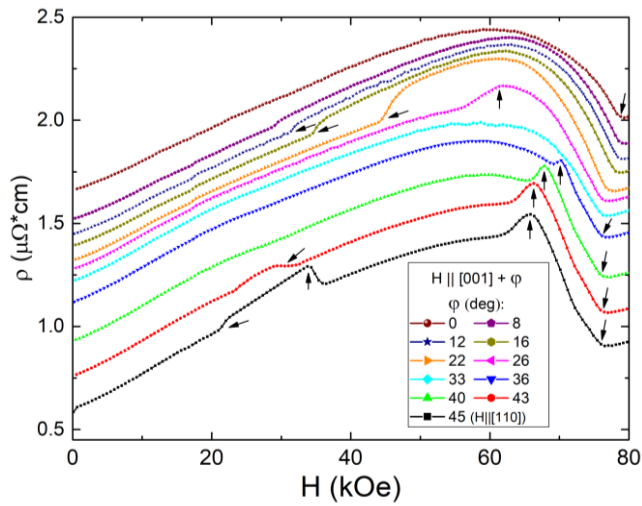

**Fig.4.** Magnetic field dependencies of the Ho$^{11}$B$_{12}$ resistivity at $T = 2.1$ K for different angles $\varphi = (n, H)$ between the normal to the sample $n$ and magnetic field $H$. Arrows show

orientational magnetic phase transitions. The rotation was performed around the axis $I\|[100]$. Curves are shifted for convenience.

**Derivatives dρ/dH**

To detect the positions of phase boundaries on the H-φ phase diagram from the field dependencies of the resistivity, numerical differentiation was fulfilled and the derivatives dρ/dH were constructed (fig.3). Numerous magnetic field transitions (marked as arrows) are observed. It is worth noting the presence of an orientation phase transition in the range of magnetic fields 15–20 kOe (shown in the inset in Fig. 3), which can be distinguished only in derivatives and is hardly visible in the initial resistivity curves.

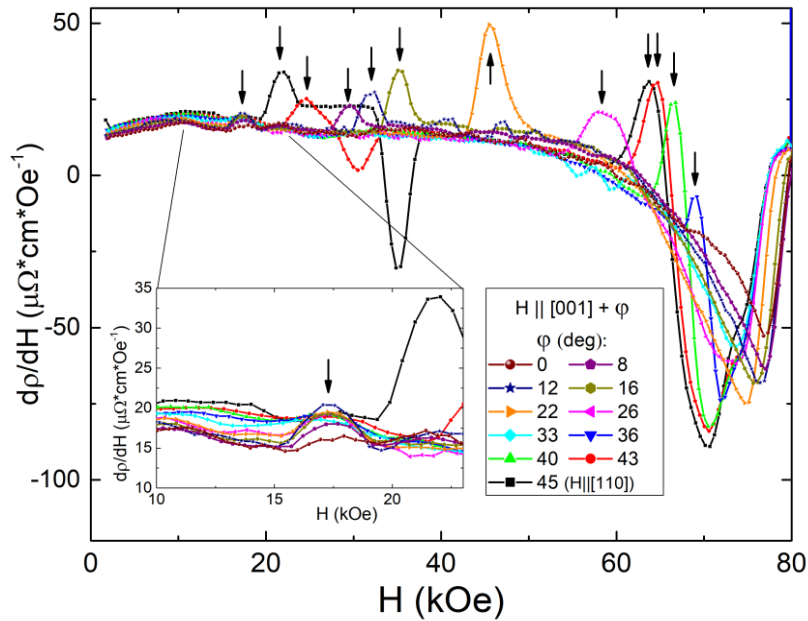

Fig.5. Magnetic field dependencies of dρ/dH derivatives for different angles φ=(**n**, **H**) between the normal to the sample **n** and the external magnetic field **H**. Arrows show orientation magnetic phase transitions. The range of 10-23 kOe is shown in zoom on the inset. The rotation was performed around the axis $\mathbf{I}\|[100]$.

**Rotation around $\mathbf{I}\|[111]$ (iii) and $\mathbf{I}\|[112]$ (iv)**

As mentioned in the article, conducting a series of experiments with the sample rotation around the measuring current axis, we obtain a series of planes in the phase H-φ-θ space that is perpendicular to the current direction. The spatial form of various phases can be reconstructed using the boundaries of the obtained H-φ diagrams. Moreover, for a more accurate determination of the shape and boundaries of the phases thus obtained, it seems essential to conduct more experiments with different current directions. Assuming this, we additionally oriented and prepared for measurement samples with current directions $\mathbf{I}\|[111]$ and $\mathbf{I}\|[112]$. The data obtained from experiments with these samples were processed in the same way as in the article and presented in Fig. 4 (a) and (b) for $\mathbf{I}\|[111]$ and $\mathbf{I}\|[112]$ respectively.

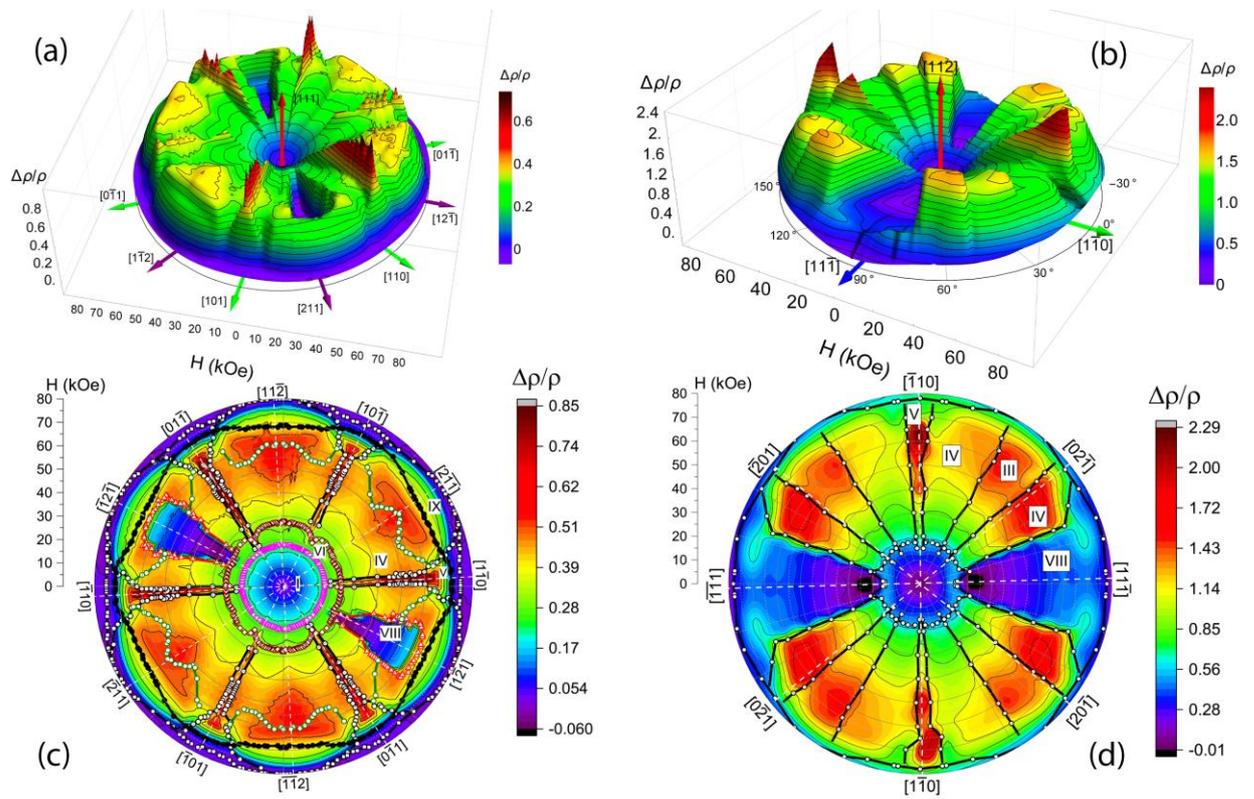

Fig.6 Magnetoresistance $\Delta\rho/\rho = f(H,\varphi)$ of the Ho$^{11}$B$_{12}$ in cylindrical coordinates for the current direction I||[111] (a) and I||[112] (b) and its projection onto (111) and (112) surfaces (c) and (d) respectively at T=2.1K. Roman numerals show different magnetic phases in the AF state.

[1]     H. Werheit, V. Filipov, K. Shirai, H. Dekura, N. Shitsevalova, U. Schwarz, and M. Armbruster, Raman scattering and isotopic phonon effects in dodecaborides, J. Phys.: Condens. Matter **23**, 065403 (2011). 10.1088/0953-8984/23/6/065403